\documentclass[conference]{IEEEtran}
\usepackage[utf8]{inputenc}
\usepackage{graphicx}
\usepackage{float}
\usepackage{mathtools}
\usepackage{subfig}
\usepackage[table,xcdraw]{xcolor}
\usepackage{lettrine}
\usepackage{nopageno}
\usepackage{algorithm,algorithmic}
\usepackage{url}
\usepackage{cite}
\usepackage{multirow}
\usepackage{tikz}
\usepackage{siunitx}
\usepackage[justification=centering, font=normalsize]{caption}
\title{Energy and Service-priority aware Trajectory Design for UAV-BSs using Double Q-Learning} 

 \author{\IEEEauthorblockN{Sayed Amir Hoseini$^1$, Ayub Bokani$^2$, Jahan Hassan$^3$,  Shavbo Salehi$^4$ Salil S. Kanhere$^5$}
\IEEEauthorblockA{$^{1,2,3}\, $School of Engineering and Technology,
The Central Queensland University,
Sydney, Australia \\
\{s.hoseini, a.bokani, j.hassan\}@cqu.edu.au}
$^{4}\,$Electrical and Computer Engineering Department, Urmia University, Urmia, Iran\\
\ shavbo.salehi@urmia.ac.ir\\
$^{5}\, $School of Computer Science and Engineering, 
The University of New South Wales,
Sydney, Australia \\
\ salil.kanhere@unsw.edu.au
}

\begin{document}
\maketitle
\begin{abstract}
 Next generation mobile networks have proposed the integration of Unmanned Aerial Vehicles (UAVs) as aerial base stations (UAV-BS) to serve ground nodes. Despite having advantages of using UAV-BSs, their dependence on the on-board, limited-capacity battery hinders their service continuity. Shorter trajectories can save flying energy, however UAV-BSs must also serve nodes based on their service priority since nodes' service requirements are not always the same. 
% 31 Aug Shorter trajectories can save flying energy, however service priority of nodes must also be addressed by the UAV-BSs since  nodes' service requirements are not always the same. 
In this paper, we present an energy-efficient trajectory optimization for a UAV assisted IoT system in which the UAV-BS considers the IoT nodes’ service priorities in making its movement decisions. We solve the trajectory optimization problem using Double Q-Learning algorithm. Simulation results reveal that the Q-Learning based optimized
trajectory outperforms a benchmark algorithm, namely Greedily-served algorithm, in terms of reducing the average energy consumption of the UAV-BS as well as the service delay for high priority nodes. 

\end{abstract}

\section{Introduction}

Next generation wireless communication systems, e.g., 5G and beyond, have proposed the integration of Unmanned Aerial Vehicles (UAVs) as aerial base stations (UAV-BS) to provide terrestrial communication services \cite{8641419}. %%JH: Due to their dynamic deployment and autonomous flight control features, UAV-BSs have various benefits over terrestrial base stations, such as on-demand coverage provisioning in areas of need (e.g., disaster area) and enhancing spectral efficiency as well as user Quality-of-Experience (QoE). 
Due to their dynamic deployability and mobility, UAVs can act as flying base stations to collect sensor data by moving closer to the Internet of Things (IoT) sensor nodes in remote farming or disaster areas without cellular or Wi-Fi coverage. An implementation of such a UAV-based sensor data collection system in the rural area of China for agricultural monitoring has been reported in ~\cite{dronesInAgri_2018}.
However, unlike the terrestrial base stations that have continuous power supply, the UAV-BSs, or UAVs in general, have limited power supply from their on-board battery. For example, off-the-shelf drones such as DJI Spreading Wings S900 can stay afloat for around 18 minutes when fully charged \cite{Azade_Survey}. Once the power is drained, the service provided by the UAV needs to pause for battery replacement or recharging. This leads to service-time limitations and discontinuity of UAV-delivered services.

\begin{figure}[h]
\centering
\includegraphics[width=0.47\textwidth]{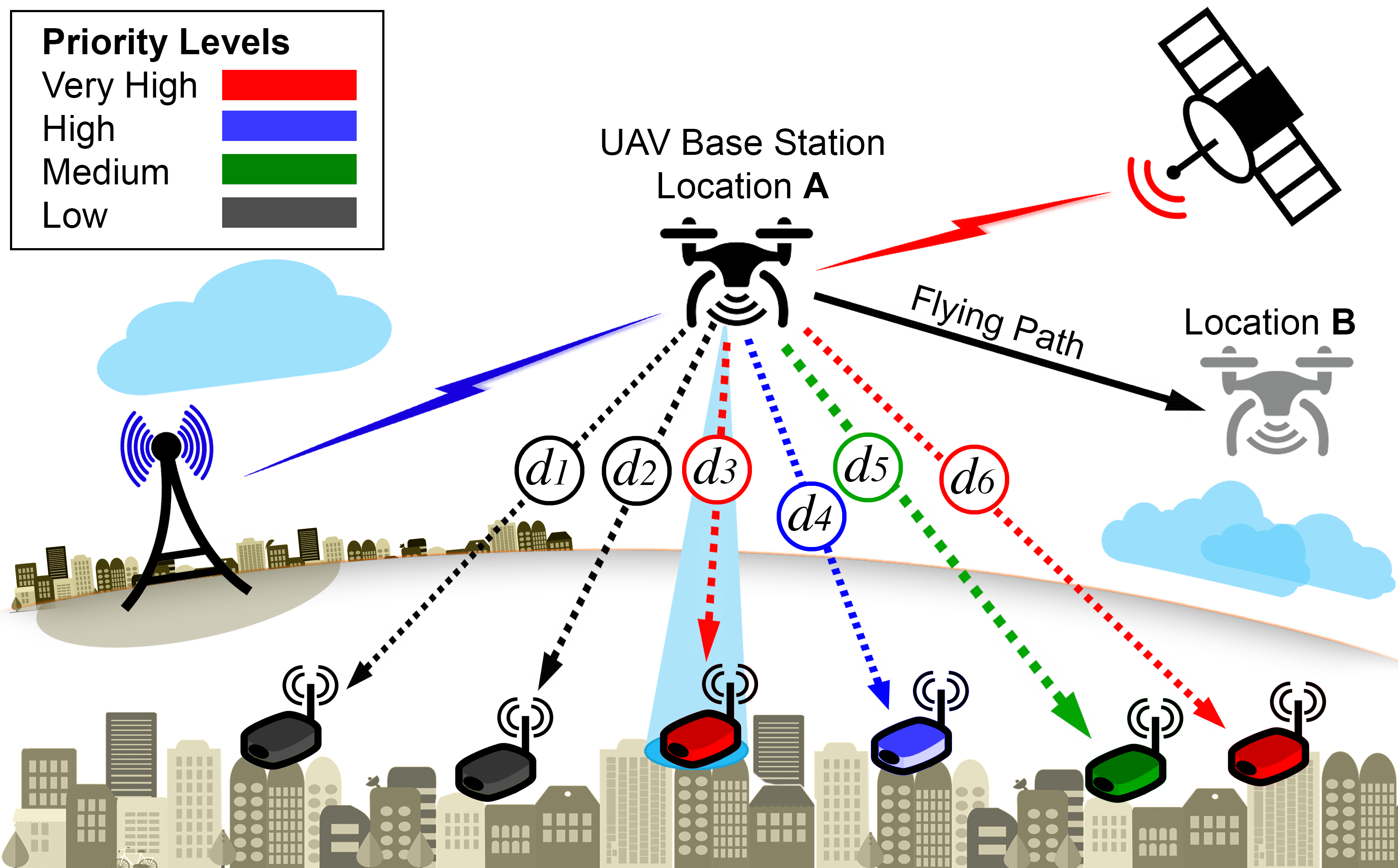}
\caption{\textbf{UAV assisted IoT System} where the UAV serves nodes, e.g., IoT devices or sensors, with different priority levels (shown with color-coding) and distances (shown with labelled arrows). The nodes should be in communication range, therefore to achieve this the UAV will fly above the nodes at a certain distance to collect data.}
\label{fig:sys-mod}
\end{figure}
%%the power limitation issue certainly motivates the design of methods and mechanisms for UAVs to function in an energy-efficient manner so that the batteries can last longer irrespective of the charging mechanisms.
Although some researchers are looking into aerial recharging of UAVs to address this \cite{wcnc_hassan, infocom_recharging}, the power limitation issue certainly motivates the design of methods and mechanisms for UAVs to function in an energy-efficient manner allowing the UAVs to do more within their battery budget, irrespective of the charging mechanisms. Since the major energy consumption of an UAV comes from its mechanical actions, e.g., flying, shortening the flight length in its mission is an effective way to reduce energy consumption, which inevitably leads to a trajectory optimization problem. While conserving energy is highly important, in a UAV-based IoT sensor data collection scenario, the UAV must also serve ground sensors as per their priority or urgency levels, e.g., based on their delay tolerance \cite{priority_UAV} levels. Therefore, while optimizing the trajectory for energy-savings, the nodes' priorities must be considered by the UAVs. Related works in trajectory design address these two requirements independently; as examples, see \cite{zeng2019energy} for energy-efficient trajectory optimization, and \cite{priority_UAV, priority-sensorUAV} for IoT nodes' priority-based trajectory optimization. %We provide more details on related work in the next section. 
Our aim is to combine both requirements in the trajectory design. 
%%%JH 25 AUG-- this is original text, above I am changing it a bit
%%JH 25 Aug --While conserving energy is highly important, UAV-BSs must also support ground nodes as per their service priority based on QoS requirements \cite{8579209}, or urgency levels, e.g., as in a UAV assisted Internet of Things (IoT) system for data collection \cite{priority-sensorUAV}. Therefore, while optimizing the trajectory for energy-savings, the nodes' request priorities must be considered by the UAVs. Related works in trajectory design of UAVs address these two requirements independently; as examples, see \cite{zeng2019energy} for energy-efficient trajectory optimization, and \cite{priority_UAV} for heterogeneous IoT nodes' priority-based trajectory optimization. We provide more details on related work in the next section. Our aim is to combine both requirements in the trajectory design. 

In this paper, therefore, we design a UAV-BS's trajectory in a UAV assisted node priority oriented IoT system that minimizes flying costs while serving nodes as per their  priorities.
%This needs to consider all effective parameters that can be available as different state components to make the movement decision by the UAV-BS prior to serving the next node.
Therefore, an intelligent model that the UAV-BS can use to make the best node-visiting decisions at different states is required. The problem can be formulated as that of selecting an action from a finite set of choices based on the repetitive observations of the environment. We optimize the UAV path using Double Q-Learning \cite{hasselt2010double} which is a model-free reinforcement learning algorithm. Q-Learning not only learns in which order the nodes should be served by the UAV-BS after some experiences, but also can dynamically update the decision policy if the environment or nodes behaviour changes. Our contributions in this paper are summarized below:

\begin{itemize}
   % \item We propose the problem of an energy-efficient trajectory design for UAV-BSs while considering nodes' service priorities. 
    % \item We show that the energy-efficient, service-priority aware trajectory optimization of the UAV-BS can be modeled as a Markov Decision Process (MDP) given that the movement decision of the UAV-BS at any  given  step  affects the long-term  discounted  utility  of  the underlying IoT system supported by it.  
    
    \item We demonstrate that the UAV-BS trajectory can be optimized based on energy-efficiency and service-priority using Double Q-Learning in a UAV-based sensor data collection scenario, and evaluate the  performance  of  the  proposed  system using  simulations.  
\item Our  simulation results  confirm  that  the optimized  trajectory  not only achieves significant energy savings compared to the bench-marking algorithm, but also serves nodes as per their priorities thereby enhancing the practicality of such systems. 
\end{itemize}

 The rest of the paper is organized as follows: Section \ref{sec:relwork} describes related work on both energy-efficient trajectory design and priority-based services in UAV-assisted networks. In Section \ref{sec:mdp}, we present our system model and Double Q-Learning formulation of the optimization problem. Section \ref{sec:perf} provides details of our simulation studies and a discussion of results. Finally, we conclude our paper in Section \ref{sec:conclu}.

\section{Related Works}
\label{sec:relwork}
%\subsection{Energy-Efficiency in Trajectory Design}

%Energy consumption in UAVs come from both mechanical activities (e.g., flying) as well as electronic activities (e.g., wireless communication), however,
The electronic energy consumption in UAVs is negligible compared to the total energy consumption \cite{drone_energy}. As such, related work predominantly focuses on the reduction of mechanical energy consumption in various ways, such as by controlling flight radius, speed, and height \cite{Azade_Survey}. For UAV-BSs, another way to reduce the mechanical energy consumption is by optimizing the path planning or trajectory design as proposed in \cite{7888557, 7101619, zeng2019energy}, etc. Zeng et al. \cite{7888557} presented an energy-efficient trajectory optimization for UAVs which also considered the communication throughput as a performance metric. 
%In this study, the propulsion energy consumption of the fixed-wing UAV at a fixed altitude was considered in the trajectory optimization, and a theoretical model was derived. 
Authors of \cite{zeng2019energy} addressed the total energy minimization of wireless communication with rotary-wing UAV by optimizing the propulsion energy as well as the communication energy in the trajectory design. Work by authors in \cite{power-constrained_traj} focused on the trajectory optimization of a rotary-wing UAV, acting as a relay node in a cellular network setting, whereby they aim to achieve a trade-off between long term communication delay and power consumption for UAV's mobility.

%\subsection{Priority-based Services in UAV-assisted IoT Networks}
Authors in \cite{priority-sensorUAV} presented a data acquisition framework for sensor networks using an UAV to collect sensor data, whereby authors introduce sensor-node priorities in the frame selection at the MAC layer. The sensor nodes get assigned different transmission priorities in different frames, based on their location with respect to the UAV, resulting in throughput maximization. Wang et al. \cite{priority_UAV} presented a trajectory design for UAV-based, time-sensitive IoT network where the nodes are assigned various priorities based on their delay tolerance sensitivity. Authors used these priorities in the cost function of a Deep Q-Learning based optimization of the trajectory to minimize system cost which they defined in terms of latency performance of the heterogeneous network.

IoT nodes' priority level values can also be set \textit{using nodes' residual energy levels} \cite{WSN_PriorityLevels} which determines active/sleep schedule of sensor nodes. This information can be used by the UAV to determine node serving priority as nodes with lower residual energy are preferred to be served first.
%served before nodes with higher residual energy since the lower energy nodes will go to sleep mode sooner than the higher energy nodes. 

In our prior work \cite{MICC_ourwork}, UAV-BS trajectory design was optimized using the Travelling Salesman Problem (TSP) method for energy efficiency but the applicability of TSP is limited when the service priory has to be considered as well. Therefore, we employ Double Q-Learning in this paper to optimize the trajectory considering both energy efficiency and service priority.

\section{System Model and Q-Learning Formulation}
\label{sec:mdp}

%We study a UAV assisted IoT system where the UAV is deployed to provide data collection services to ground sensor nodes or IoT devices, referred to as `nodes' henceforth in the paper. 
In our considered scenarios, we assume static ground nodes that are randomly positioned in different locations and require data gathering services from the the UAV with different priority levels as shown in Figure \ref{fig:sys-mod}. %The priority levels are based on the sensor nodes' delay tolerance sensitivity, similar to that used in \cite{priority_UAV}. 
Priority values change as per nodes' residual energy level. Nodes' location and initial service priority are pre-loaded in UAV software. Each time the UAV collects data from a node, it learns the updated node's residual energy level and once all nodes are served, the new values are used by the UAV to determine the priorities in the next round. Since, the UAV should take action based on the observed environment and each experience can contribute the decision enhancement, we find this optimisation is fitted to Reinforcement Learning.

We model the service area on the ground as a grid and consider the UAV as the Q-Learning agent. The state space is created by the UAV's observations of its own location, nodes' locations, and service priority level of each node. The Q-Learning agent must take action and fly to the next node which should be served. Therefore, the number of possible actions equals to the number of unserved nodes. The Q-Learning model consists of following four basic components:
\begin{itemize}
\item \textbf{Agent} (UAV) is the flying base station which should serve nodes one by one based on their locations and service priority levels.
\item \textbf{Action} (a) is the UAV's next flying destination which is determined by the location of next node to serve. 
\item \textbf{State} (S) is defined based on the observed information of UAV's current location and nodes information. Therefore, our system state is defined as $S=\{L_{uav},L_{nd},\Omega_{nd}\}$ where $L_{uav}$ is UAV's location, $L_{nd}=[L_{nd_1},L_{nd_2}, ...,L_{nd_n}]$ is a vector that represent the location of node 1 to $n$ and $\Omega_{nd} = [\Omega_{nd_1},\Omega_{nd_2},...,\Omega_{nd_{n}}]$ is a vector that denotes nodes' service priority ($sp$) and status (i.e., served or not). We consider a number from 1 to 4 to represent low, medium, high and very high $sp$ levels respectively. Once a node is served, we set its $sp$ level to zero to distinguish it from the other nodes waiting to be served. 
\item \textbf{Revenue} (R) is a function that returns a real number for each $state-action$ pair after the Q-Learning agent has moved to the next state $s'$ by taking action $a$. Our revenue function is a linear combination of $rewards$ and $penalties$. Serving a node results in a reward, whereas penalties are given for flying energy cost and service delay.
%% OUR OLD ONE: Serving each node is rewarded while flying energy cost and the service delay are penalized.  
 
\end{itemize}

In this model, we consider variable time steps due to having different distances between the UAV and each ground node. At each time step, the UAV selects the next waiting node and flies towards it. The service time is assumed to be negligible and the UAV is assumed to fly at a fixed speed and quickly communicates with nodes. To save ground nodes' energy, the UAV communicates data in the closest distance with these devices which means the UAV collects data when it is on top of them. In the Q-Learning the rewards of each state-action are saved in Q-Table and are being updated by new experiment. While traditional Q-Learning uses one Q-Table, the double Q-Learning employs two Q-Tables to avoid possible local optima and achieves the global optimum. Hence, we denote these Q-Tables as $Q_A$-Table and $Q_B$-Table. After serving each node, the Q-values of the Q-table that was used in serving the node are updated using the relevant double Q-Learning equation from the below equations:
\begin{equation}
      Q_A^{new}(s,a) = \\
      (1-\alpha) Q_A(s,a) + \alpha (R(s,a) + \gamma Q_B(s',a^{*})),
    \label{eq:QL-main-A}
\end{equation} 
\begin{equation}
      Q_B^{new}(s,a) = \\
      (1-\alpha) Q_B(s,a) + \alpha (R(s,a) + \gamma Q_A(s',b^{*})),
    \label{eq:QL-main-B}
\end{equation} 
where $\alpha$ and $\gamma$ are the learning rate and discount factor respectively. $R$ is the revenue function, s$'$ is the next state after taking action a at state s and $a^{*}$ and $b^{*}$ are the maximum Q-value of all state-action pairs on state s$'$ as:
\begin{equation}
    a^{*}=\underset{a}{\mathrm{arg\,max}}\; Q_A(s',a),
    \label{eq:bestActiona}
\end{equation}
\begin{equation}
    b^{*}=\underset{a}{\mathrm{arg\,max}}\; Q_B(s',a).
    \label{eq:bestActionb}
\end{equation}
For the Q-Learning revenue function $R(s,a)$, we consider a \textit{reward} when the UAV delivers services with a high priority and we apply different $penalties$ for the service delivery delays and flying energy consumption.
%The delay penalty is proportional to the node's service priority. 
Such considered rewards and penalties help the Q-Learning agent, i.e., the UAV, to learn the model and find an optimal trajectory that minimizes the total energy consumption, serves the least delay tolerant nodes first, and increases the overall QoE. Hence, The revenue function R is calculated as:
\begin{equation}
  R = w_1\Omega_{nd_a} - w_2\sum_{i=1, i\neq a}^{n}\Omega_{nd_i} t_s -w_3\int_{0}^{t_s}P(V)dt,
  \label{eq:revenue}
\end{equation}
where $w_1$, $w_2$ and $w_3$ are the tuning parameters, $nd_{a}$ is the served node, and $t_s$ is elapsed time from serving the last node. $P(V)$ is UAV's power consumption for flying with speed $V$ which is calculated as \cite{zeng2019energy}:
%%%%%%%%%%%%%%%%%%%%%%%%%

\begin{equation}
    \begin{aligned}
     P(V)=&P_0\left ( 1+\frac{3V^2}{U^2} \right )+P_i\left ( \sqrt{1+\frac{V^4}{4\nu_{0}^4}} -\frac{V^2}{2\nu_{0}^2}\right )^{1/2}\\    &+\frac{1}{2}d_0\rho sAV^3,
    \end{aligned}
\end{equation}

where relevant parameters are listed in Table \ref{tab:sim_param}.

\begin{algorithm}[t]
\caption{: Double Q-Learning Episode}\label{alg:Q_Learning_Algorithm}
\begin{algorithmic}[]
 \footnotesize
    \STATE  \ Load $Q_A$-Table and $Q_B$-Table from previous episode
	\STATE  \ Initialize UAV location
	\STATE  \ exploration = epsilon
    \STATE  \ Observe nodes' location
    \STATE  \ Set Q-Selector = A
    \STATE  \ \textbf{while} there is a node to serve \\
    \ \textbf{do} \{
    \STATE  \ \ \ \ Observe nodes' service priority
    \STATE  \ \ \ \ Current state = (UAV location, nodes' location, nodes' priority)
    \STATE  \ \ \ \ Generate a random number $p$ $\in$[0,1]
    \STATE  \ \ \ \ \textbf{if} $p$ is less than Exploration
    \STATE  \ \ \ \ \ \ \ \  Select next node randomly 
    \STATE  \ \ \ \ \textbf{else}
    \STATE  \ \ \ \ \ \ \ \ \textbf{if} Q-Selector = A
    \STATE  \ \ \ \ \ \ \ \ \ \ \ \ Choose next node based on $Q_A$-Values
    \STATE  \ \ \ \ \ \ \ \ \textbf{else if} Q-Selector = B
    \STATE  \ \ \ \ \ \ \ \ \ \ \ \ Choose next node based on $Q_B$-Values
   % \STATE  \ \ \ \ \ \ \ \ \textbf{end if}
    %\STATE  \ \ \ \ \textbf{end if}
    \STATE  \ \ \ \ Move UAV to next node location
    \STATE  \ \ \ \ Current node = Next node
    \STATE  \ \ \ \ Collect current node data
    \STATE  \ \ \ \ Calculate Revenue(consumed energy by UAV, \\
    \ \ \ \ \ current node service priority, service delay)
    \STATE  \ \ \ \ \textbf{if} Q-Selector = A
    \STATE  \ \ \ \ \ \ \ \  Update $Q_A$-Table using (\ref{eq:QL-main-A})
    \STATE  \ \ \ \ \textbf{else if} Q-Selector = B
    \STATE  \ \ \ \ \ \ \ \  Update $Q_B$-Table using (\ref{eq:QL-main-B})
    \STATE  \ \ \ \ Set current node service priority to zero
    \STATE  \ \ \ \ Toggle Q-Selector between A and B
    \STATE \ \ \ \ \textbf{\}}
    \STATE  \ Decrease epsilon for the next episode
 %   \STATE  \ \textbf{Continue}
\end{algorithmic}
\end{algorithm}

\begin{algorithm}[b]
\caption{: Greedy}\label{alg:NN}
\begin{algorithmic}[]
 \footnotesize
	\STATE  \ Initialize UAV location
    \STATE  \ Observe nodes' location
    \STATE  \ \textbf{while} there is a node to serve \\
    \ \textbf{do \{}
    \STATE  \ \ \ \ remaining nodes = nodes with non-zero service priority
    \STATE  \ \ \ \ Calculate the distance between UAV and remaining nodes
    \STATE  \ \ \ \ Select the node with minimum distance as next node
    \STATE  \ \ \ \ Move UAV to next node location
    \STATE  \ \ \ \ current node = next node
    \STATE  \ \ \ \ Collect current node data
    \STATE  \ \ \ \ Set current node service priority to zero
    \STATE  \ \ \ \ Decrease Exploration 
    \STATE  \ \ \ \ \textbf{\}}
%    \STATE  \ \textbf{Continue}
\end{algorithmic}
\end{algorithm}

We use \emph{epsilon-greedy} scheme \cite{sutton1998reinforcement} in our algorithm where actions are taken randomly at the beginning of learning process and the agent is fully in \emph{exploration} mode. Besides, by decreasing the epsilon value,  the chance of \emph{exploitation} increases and the action with the highest Q-Value may be taken at each step. This is adjusted to rely on Double Q-Learning policy gradually over the time.

At each step of time, UAV observes the state $s$, then takes an action $a$ and receives a revenue $R(s,a)$ after moving to the state $s'$. The goal of the training phase is to find the sequential order of served nodes that maximizes the total future revenues. Our revenue function will result in finding a flying path that minimizes energy consumption and improves the QoE. Each episode of Double Q-Learning is presented in Algorithm \ref{alg:Q_Learning_Algorithm}. The $Q_A(s,a)$ and $Q_B(s,a)$ values are saved and updated in $Q_A$-Table and $Q_B$-Table. The algorithm forces the UAV to choose  between \emph{exploration} or \emph{exploitation} approach for each decision. In exploration, UAV selects the next node to serve randomly and in exploitation, UAV takes the action with the highest Q-value for the observed state in one of the Q-Tables in turn. The exploration rate is adjusted by $\epsilon$ which is set to 1 in the early episodes to keep actions fully random and boost the training. Then through the episodes, it is decreased granularly to zero or a small value when the Double Q-Learning policy is reliable enough and most of the actions are taken based on Q-Values.      

To compare Double Q-Learning performance, we consider a Greedy algorithm as a baseline. The Greedy (Nearest Neighbor) is presented in Algorithm \ref{alg:NN} in which the UAV selects the nearest node to serve at each step, whereas the Double Q-Learning tries to achieve a balance between distance and node priority.

\begin{figure}[]
\centering
    \subfloat[Accumulated reward]{\label{fig:accuReward1}
\includegraphics[width=0.5\textwidth,clip=true, trim=30 0 0 0]{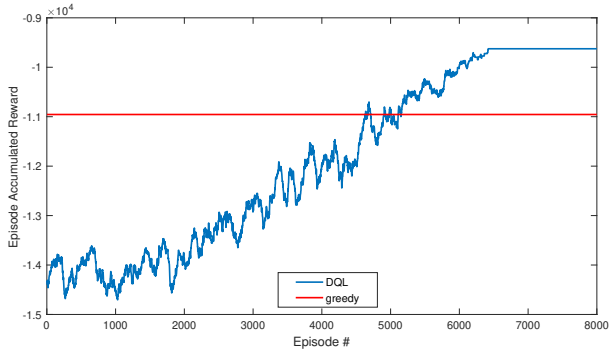}}\vspace{0em}
        \subfloat[Total UAV energy consumption on the episode]{\label{fig:energy1}
\includegraphics[width=0.5\textwidth,clip=true, trim=30 0 0 0]{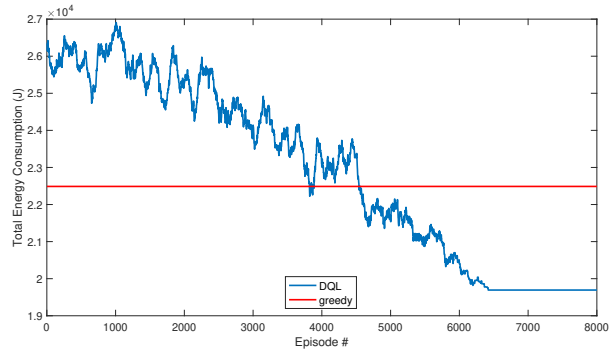}}
    \caption{Double Q-Learning performance vs. Greedy (Nearest Neighbor) for scenario \#1.}
    \label{fig:sc1-learningResults}
\end{figure}

\section{Performance Evaluation}
\label{sec:perf}

\subsection{Simulation Setup}

\begin{figure*}
    \centering
    \subfloat[Greedy Algorithm]{\label{fig:trajectory-1NN}
\includegraphics[width=0.39\textwidth]{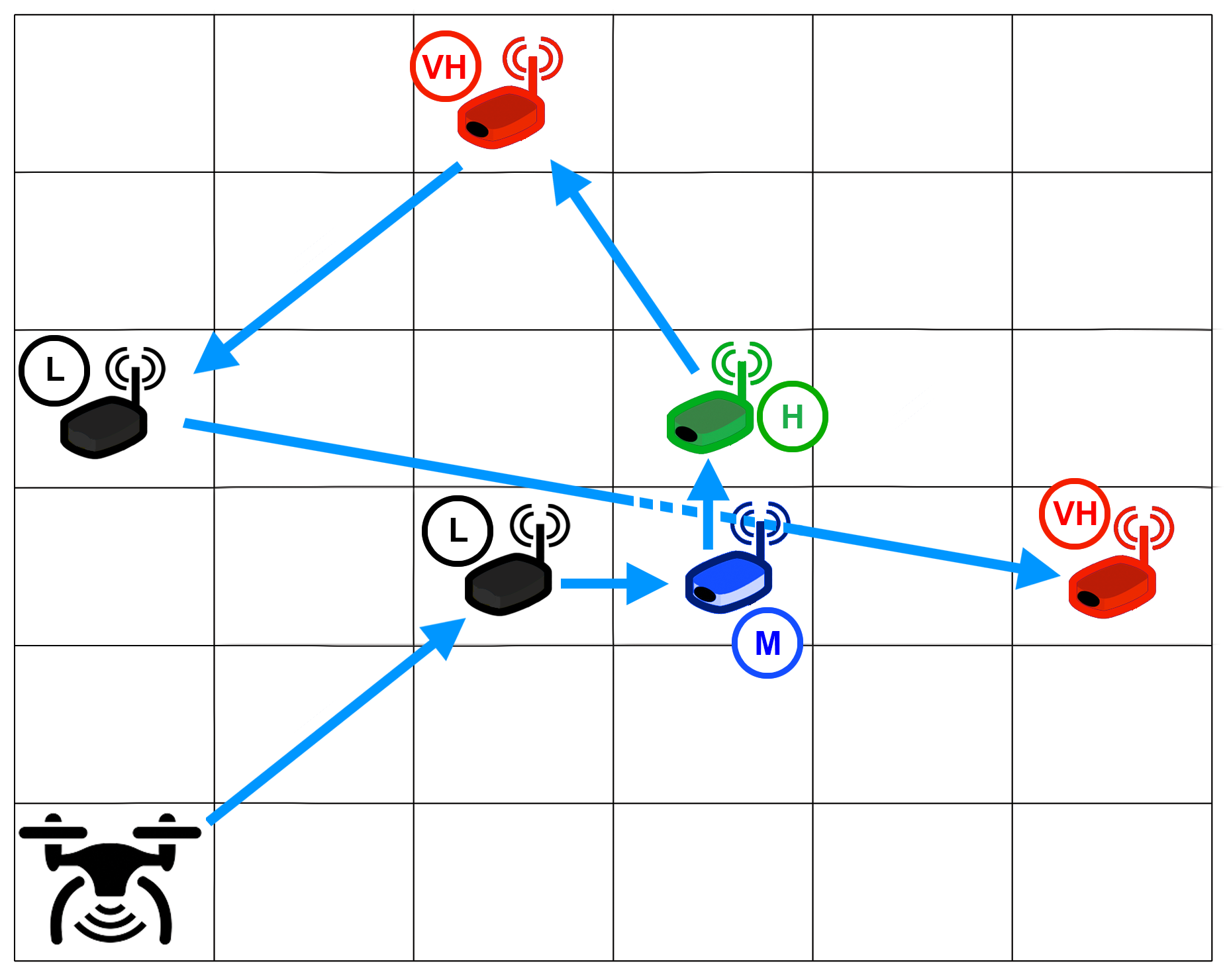}}
        \subfloat[Double Q-Learning]{\label{fig:trajectory-1QL}
\includegraphics[width=0.39\textwidth]{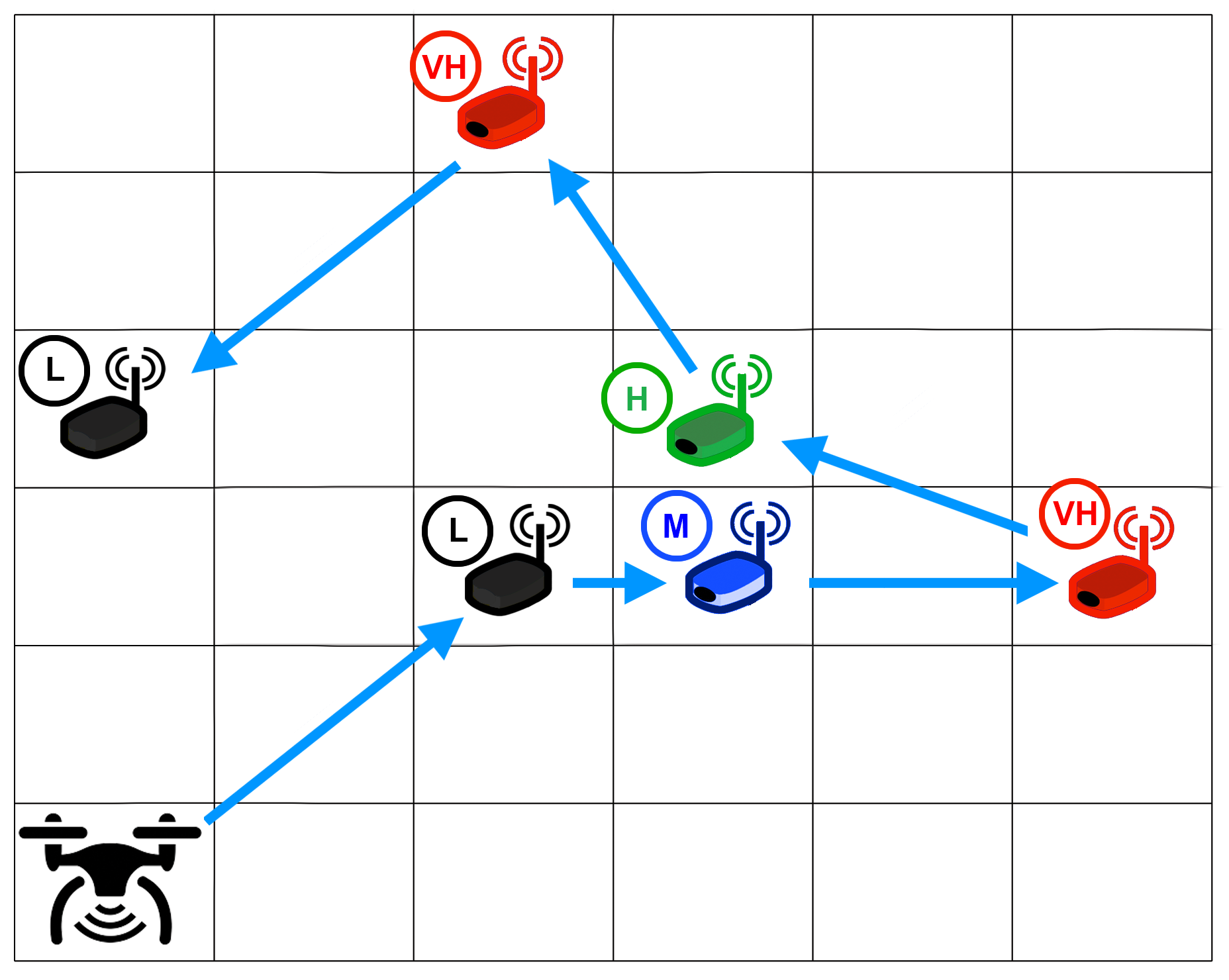}}
    \caption{UAV base station trajectory for scenario \#1.}
    \label{fig:trajectory-1}
\end{figure*}

\begin{figure*}[]
    \centering
    \subfloat[Greedy algorithm]{\label{fig:sc2-NN}
\includegraphics[width=0.4\textwidth]{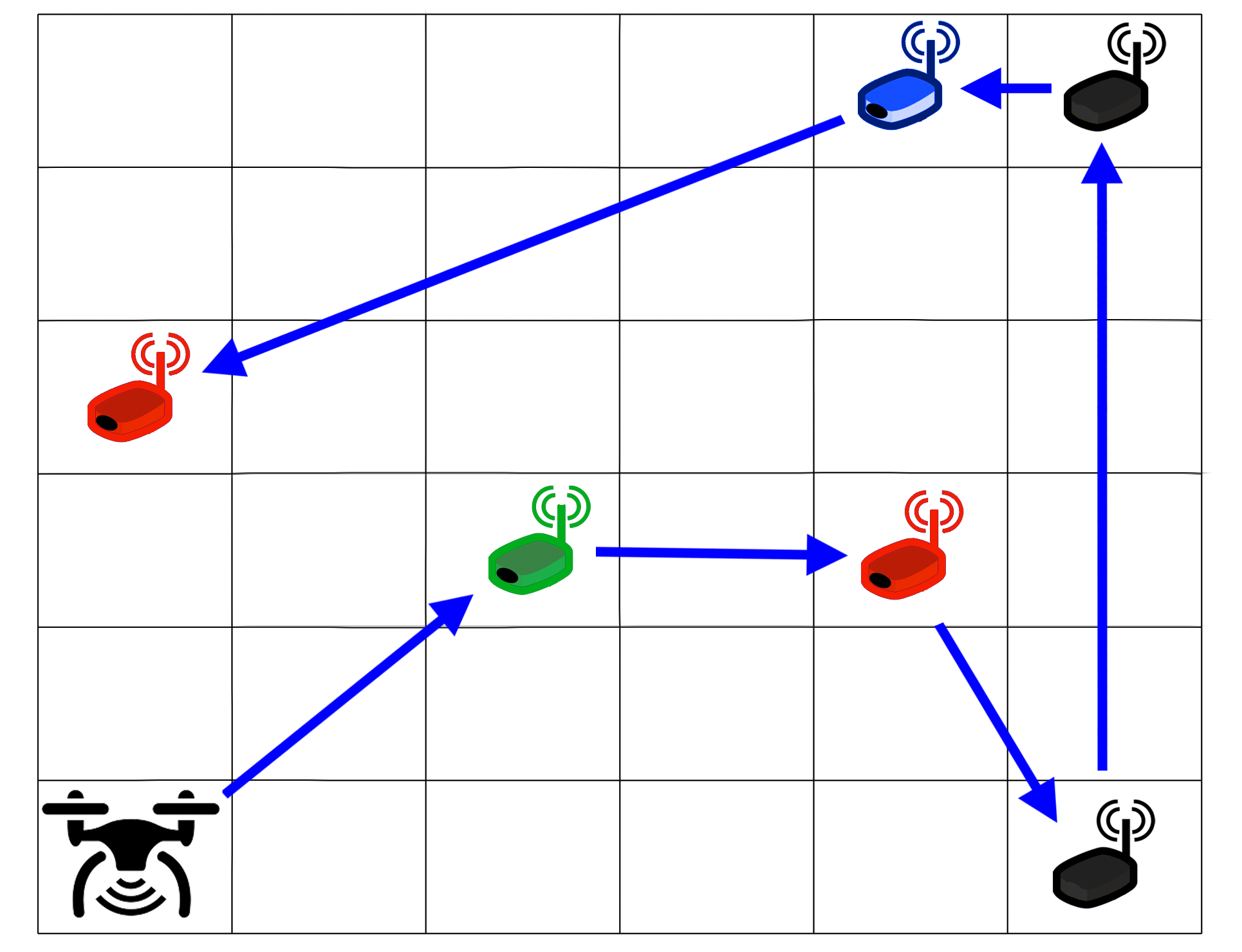}}
    \subfloat[DQL1: High revenue for service to high priority node. $w_1=30000$, $w_2=7.5$, $w_3=0.1$. ]{\label{fig:sc2-srw}
\includegraphics[width=0.4\textwidth]{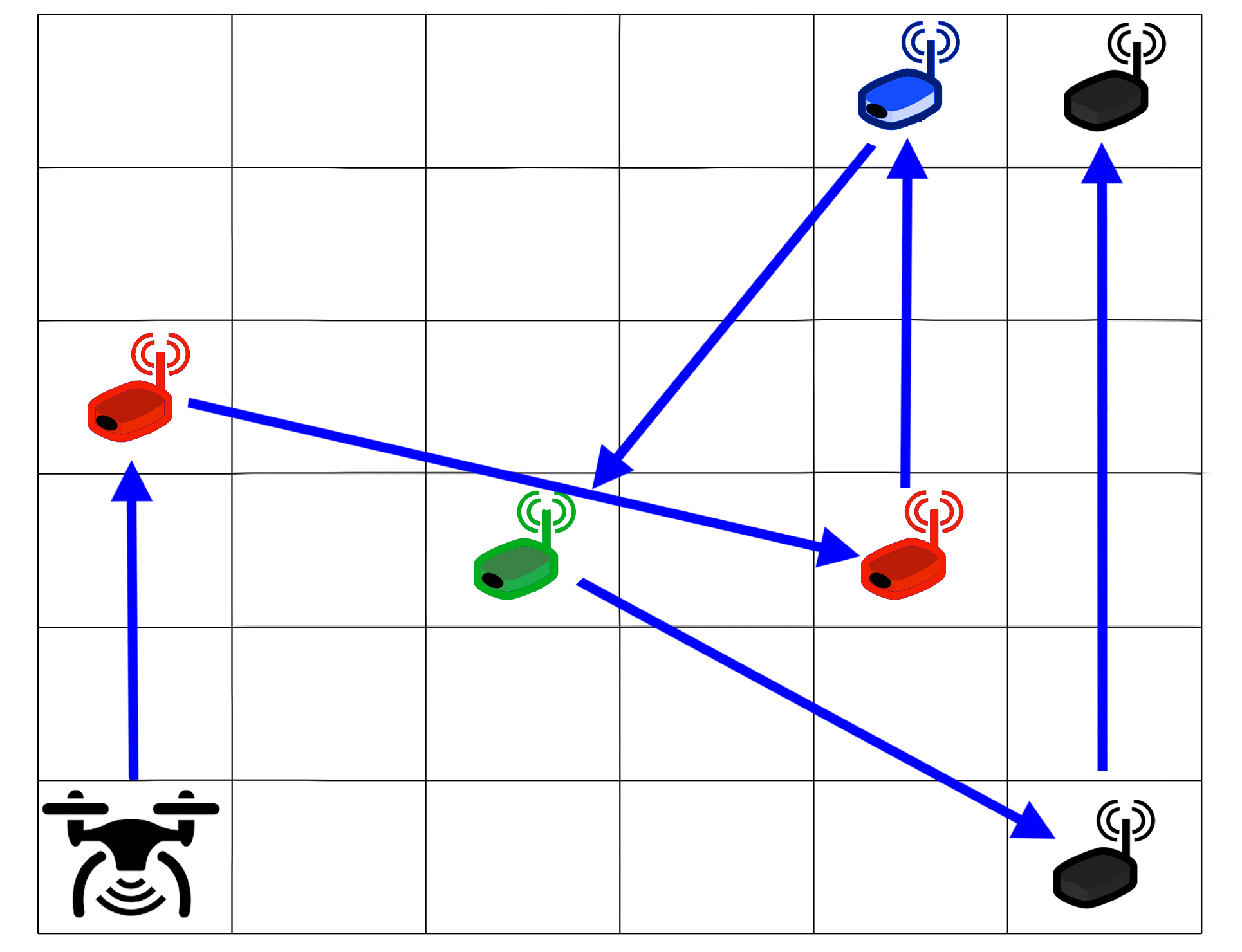}}\vspace{0em}
        \subfloat[DQL2: High revenue for QoE.$w_1=30$, $w_2=750$, $w_3=0.1$]{\label{fig:sc2-QL-dcw}
\includegraphics[width=0.4\textwidth]{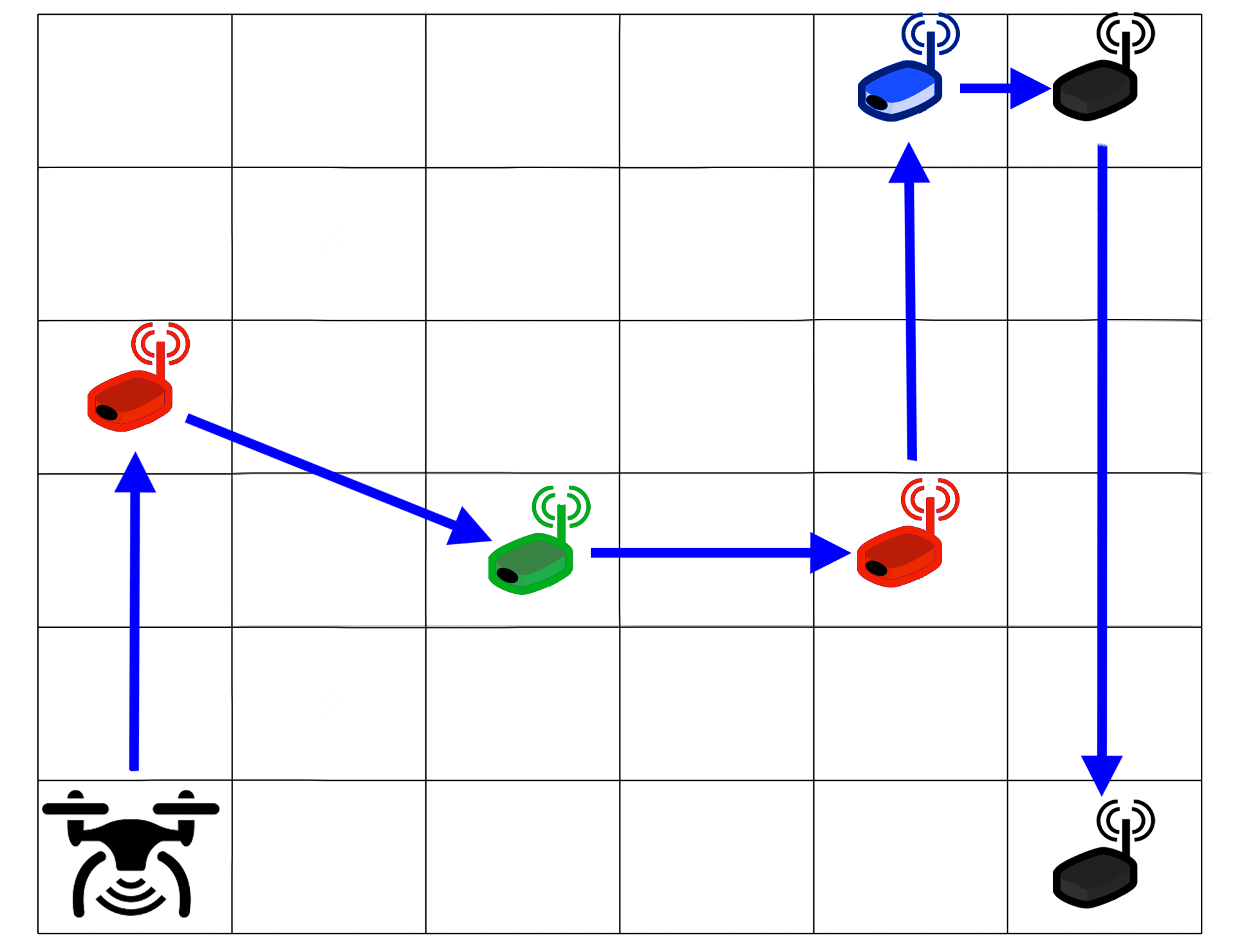}} 
    \subfloat[DQL3: High penalty for energy consumption. $w_1=30$, $w_2=7.5$, $w_3=100$.]{\label{fig:sc2-ecw}
\includegraphics[width=0.4\textwidth]{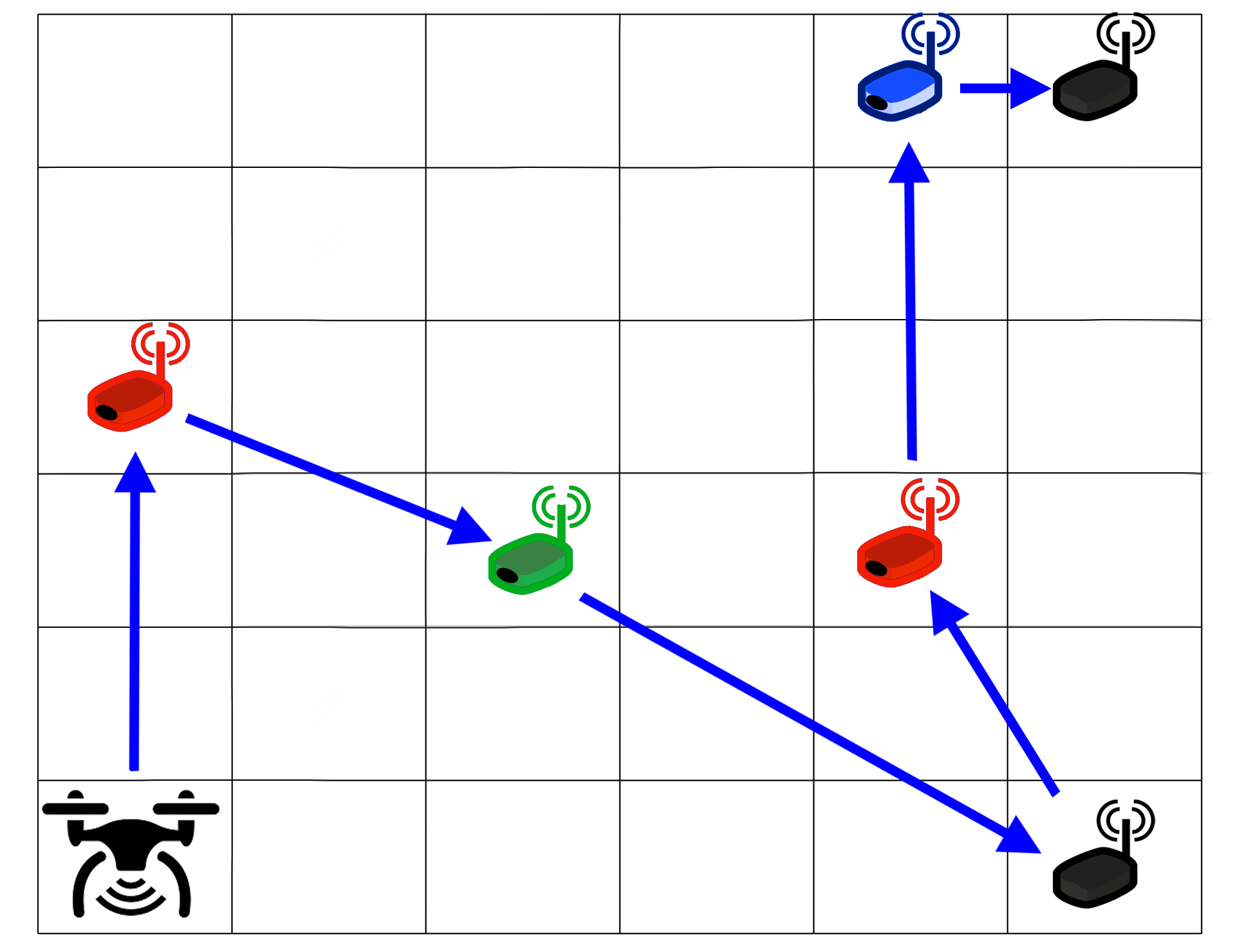}}

    \caption{Greedy and Double Q-Learning Trajectories for scenario \#2. Double Q-Learning results are generated for different revenue adjustment parameters.}
    \label{fig:trajectory-2}
\end{figure*}

To evaluate the proposed method, we simulated two scenarios as illustrated in Figures ~\ref{fig:trajectory-1} and \ref{fig:trajectory-2} respectively, where six nodes are randomly distributed on a 6 by 6 grid service area. Each node has data for transmission with a random service priority which is represented by different colours as was illustrated in Figure \ref{fig:sys-mod}. As mentioned above, $\epsilon$-greedy scheme is used to force Double Q-Learning to take random actions at the start which is helpful in training to avoid local convergence. $\epsilon$ is equal to 1 for the first 1000 episodes, then it is reduced slowly to zero at episode \# 6400. Therefore, the Double Q-Learning policy is fully operational for the last 1600 episodes. Simulation parameters are shown in Table \ref{tab:sim_param}.

\begin{table}[t]
\caption{Simulation Parameters}
    \centering
   \scriptsize
    \begin{tabular}{|l|l|}
\hline
 Cell side & 50 \si{m} \\ 
 Learning rate ($\alpha$) & 0.5 \\ 
 Discount factor ($\gamma$) & 0.95 \\ 
 %exploration schedule & 1 \\
 Revenue tuning parameter ($w_1$) & 30 \\ 
 Revenue tuning parameter ($w_2$) & 7.5 \si{s^{-1}}\\
 Revenue tuning parameter ($w_3$) & 0.1 \si{Joule^{-1}}\\ 
 Air density ($\rho$) & 1.225   \si{kg/{m^3}}\\ 
 Rotor disc area ($A$) & 0.181 \si{m^2}\\
 Tip speed of the rotor blade ($U$) & 96 \si{m/s}\\
 Fuselage drag ratio ($d_0$) & 0.9 \\
 Rotor solidity ($s$) & 0.1 \\
 UAV speed ($V$) & 5 \si{m/s} \\
 Blade profile power ($P_0$) & 29.4 \si{W}\\
 induced power ($P_i$) & 206.5 \si{W}\\
 Mean rotor induced velocity ($\nu _0$) & 7.5 \si{m/s}\\
 \hline
    \end{tabular}
    %\caption{Simulation Parameters}
    \label{tab:sim_param}
\end{table}

\subsection{Results}

\begin{figure}[]
\centering
\includegraphics[width=0.45\textwidth]{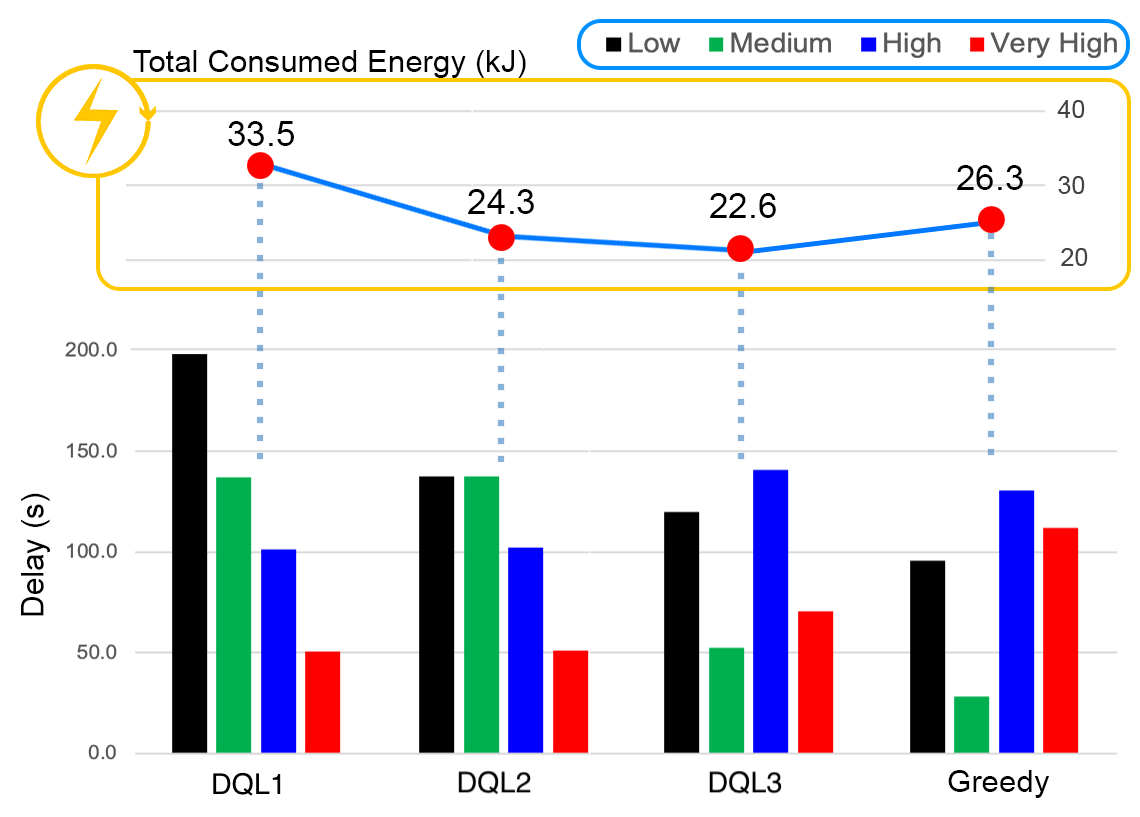}
\caption{Average Serving Delay and Energy Consumption Comparison of various revenue parameter set of Double Q-Learning and Greedy algorithm for scenario \#2.}
\label{fig:sc2-delay}
\end{figure}

The first scenario is simulated using default tuning parameters of revenue function for Double Q-Learning and Greedy algorithm. The learning process of Double Q-Learning is presented in Figure \ref{fig:sc1-learningResults} along with the Greedy algorithm result as baseline comparison. We plugged in the revenue function to the Greedy algorithm to compare accumulated reward for both algorithms. After around 5000 episodes, Double Q-Learning outperforms Greedy algorithm for both revenue and energy consumption. Moreover, the Double Q-Learning algorithm enhances QoE by serving high priority nodes first, whereas the Greedy algorithm ignores the priority values in making movement decisions. This can be observed in Figure \ref{fig:sc2-delay} which we discuss later. Figure \ref{fig:trajectory-1} shows the resulting trajectory for the same scenario for both of the algorithms, using the same priority color coding for nodes as was shown in Figure \ref{fig:sys-mod}. The UAV starts flying from the bottom left corner and collects nodes' data one by one. Note that a near-optimal trajectory policy can be obtained offline by Double Q-Learning and utilized from the beginning of flight missions and gets updated by experiencing real-world operations continuously.

The revenue function in \eqref{eq:revenue} has an essential role in Double Q-Learning behavior. The trajectory optimisation of UAV in our model depends to service priority of nodes, service delay and energy consumption of UAV. Therefore, there is a trade-off which can be adjusted by the revenue function. To explore this, we simulated the second scenario for three different tuning parameter set of the revenue function. In each set, we choose a very large value for one of $w_1$, $w_2$ and $w_3$. We also simulate the Greedy algorithm for the same scenario for comparison, with results presented in Figure  \ref{fig:trajectory-2} and Figure \ref{fig:sc2-delay}. The trajectory is plotted and the average delay is calculated after Double Q-Learning converges to a stable outcome for the fixed scenario.  While the Greedy algorithm selects node service order based on the nearest neighbor approach, Double Q-Learning results in a different trajectory depending the parameter adjustments. When $w_1$ is set to very high value as in Figure \ref{fig:sc2-srw} (DQL1), UAV collects data of high priority nodes first. This is changed to a balanced behavior when a very high $w_2$ is set as in Figure \ref{fig:sc2-QL-dcw} (DQL2). As shown in Figure \ref{fig:sc2-delay}, the balanced Double Q-Learning presents an overall better QoE and energy consumption. While serving high priority nodes results in 33.5 \si{kJ} of energy consumption, this is reduced to 24.3 \si{kJ} for the balanced mode. Further, the delay for serving low and moderate priority nodes are also reduced for the balanced Double Q-Learning in comparison to serving high priority nodes settings. Finally, we set $w_3$ to a very high value to minimize energy consumption. The result is presented in Figure \ref{fig:sc2-ecw} (DQL3). As can be seen in Figure \ref{fig:sc2-delay}, UAV is forced to travel the near shortest path and consumes only 22.6 \si{kJ} of energy which comes at a cost of high experienced delay for the higher priority nodes.

\section{Conclusion}
\label{sec:conclu}
In this paper, we studied the scenario of a UAV providing data collection services to nodes with various service priorities in a UAV assisted IoT system. We optimized the UAV's trajectory using Double Q-Learning, with a view to reduce energy consumption while serving requesting nodes as per their required service priority. Simulation results revealed that the Q-Learning based trajectory outperformed the benchmark node-serving algorithm in reducing the average energy consumption of the UAV as well as enhancing QoE in regards to the service delay for high priority nodes. Through adjusting the tuning parameter set of the revenue function, we explored various behaviour of the Q-Learning model. We found that for a balanced setting of the parameters, the Q-Learning based trajectory was able to achieve the best trade-offs in terms of energy consumption and QoE for serving the highest priority nodes while also improving the QoE for other priority-category nodes. We note that Double Q-Learning has performance limitations when applied to an extended UAV observation space, e.g., when the number of nodes increase. For a more scalable model, in our future work we aim to use Deep Reinforcement Learning (DRL), where we can employ the Deep Neural Network to estimate Q-Values in a large Q-Table. However, it is worth avoiding the complexity and instability of DRL \cite{anschel2017averaged} for small scale applications for which Double Q-Learning's efficiency is proven in this work.

\section*{Acknowledgment}
This work is supported by the Central Queensland University Research Grant RSH5137.

\bibliographystyle{IEEEtran}
\bibliography{CCNCv1}

\begin{thebibliography}{10}
\providecommand{\url}[1]{#1}
\csname url@samestyle\endcsname
\providecommand{\newblock}{\relax}
\providecommand{\bibinfo}[2]{#2}
\providecommand{\BIBentrySTDinterwordspacing}{\spaceskip=0pt\relax}
\providecommand{\BIBentryALTinterwordstretchfactor}{4}
\providecommand{\BIBentryALTinterwordspacing}{\spaceskip=\fontdimen2\font plus
\BIBentryALTinterwordstretchfactor\fontdimen3\font minus
  \fontdimen4\font\relax}
\providecommand{\BIBforeignlanguage}[2]{{%
\expandafter\ifx\csname l@#1\endcsname\relax
\typeout{** WARNING: IEEEtran.bst: No hyphenation pattern has been}%
\typeout{** loaded for the language `#1'. Using the pattern for}%
\typeout{** the default language instead.}%
\else
\language=\csname l@#1\endcsname
\fi
#2}}
\providecommand{\BIBdecl}{\relax}
\BIBdecl

\bibitem{8641419}
Y.~{Zeng}, M.~{Debbah}, D.~{Gesbert}, I.~{Guvenc}, S.~{Jin}, and J.~{Xu},
  ``{Integrating UAVs into 5G and Beyond (Guest Editorial)},'' \emph{IEEE
  Wireless Communications}, vol.~26, no.~1, pp. 10--11, February 2019.

\bibitem{dronesInAgri_2018}
G.~Sylvester, \emph{E-agriculture in action: Drones for agriculture}.\hskip 1em
  plus 0.5em minus 0.4em\relax Bangkok, Thailand: Food and Agriculture
  Organization of the United Nations and International Telecommunication Union,
  2018.

\bibitem{Azade_Survey}
A.~{Fotouhi}, H.~{Qiang}, M.~{Ding}, M.~{Hassan}, L.~G. {Giordano},
  A.~{Garcia-Rodriguez}, and J.~{Yuan}, ``Survey on uav cellular
  communications: Practical aspects, standardization advancements, regulation,
  and security challenges,'' \emph{IEEE Communications Surveys Tutorials},
  vol.~21, no.~4, pp. 3417--3442, 2019.

\bibitem{wcnc_hassan}
J.~{Hassan}, A.~{Bokani}, and S.~S. {Kanhere}, ``Recharging of flying base
  stations using airborne rf energy sources,'' in \emph{2019 IEEE Wireless
  Communications and Networking Conference Workshop (WCNCW)}, 2019, pp. 1--6.

\bibitem{infocom_recharging}
J.~Hassan, S.~Hoseini, A.~Bokani, and S.~S. Kanhere, ``Trajectory optimization
  of flying energy sources using q-learning to recharge hotspot uavs,'' in
  \emph{2020 IEEE International Conference on Computer Communications Workshops
  (INFOCOM Workshops, accepted)}, 2020.

\bibitem{priority_UAV}
N.~Wang, Y.~Xin, J.~Zheng, J.~Wang, X.~Liu, and Y.~Liu, ``Priority-oriented
  trajectory planning for uav-aided time-sensitive iot networks,'' in
  \emph{2020 IEEE International Conference on Communications Workshops (ICC
  Workshops, accepted)}, 2020.

\bibitem{zeng2019energy}
Y.~Zeng, J.~Xu, and R.~Zhang, ``Energy minimization for wireless communication
  with rotary-wing uav,'' \emph{IEEE Transactions on Wireless Communications},
  vol.~18, no.~4, pp. 2329--2345, 2019.

\bibitem{priority-sensorUAV}
S.~{Say}, H.~{Inata}, J.~{Liu}, and S.~{Shimamoto}, ``Priority-based data
  gathering framework in uav-assisted wireless sensor networks,'' \emph{IEEE
  Sensors Journal}, vol.~16, no.~14, pp. 5785--5794, 2016.

\bibitem{hasselt2010double}
H.~V. Hasselt, ``Double q-learning,'' in \emph{Advances in neural information
  processing systems}, 2010, pp. 2613--2621.

\bibitem{drone_energy}
D.~Zorbas, T.~Razafindralambo, L.~Di~Puglia~Pugliese, and F.~Guerriero,
  ``Energy efficient mobile target tracking using flying drones,'' vol.~19, 06
  2013, pp. 80--87.

\bibitem{7888557}
Y.~{Zeng} and R.~{Zhang}, ``{Energy-Efficient UAV Communication With Trajectory
  Optimization},'' \emph{IEEE Transactions on Wireless Communications},
  vol.~16, no.~6, pp. 3747--3760, June 2017.

\bibitem{7101619}
C.~D. {Franco} and G.~{Buttazzo}, ``{Energy-Aware Coverage Path Planning of
  UAVs},'' in \emph{2015 IEEE International Conference on Autonomous Robot
  Systems and Competitions}, April 2015, pp. 111--117.

\bibitem{power-constrained_traj}
M.~Bliss and N.~Michelusi, ``{Power-Constrained Trajectory Optimization for
  Wireless UAV Relays with Random Requests},'' 2020.

\bibitem{WSN_PriorityLevels}
B.~Zeng, L.~Yao, and W.~Hu, ``\BIBforeignlanguage{eng}{Priority based data
  reporting algorithm in wireless sensor networks},''
  \emph{\BIBforeignlanguage{eng}{Journal of Shanghai Jiaotong University
  (Science)}}, vol.~22, no.~1, pp. 60--65, 2017.

\bibitem{MICC_ourwork}
S.~{Salehi}, A.~{Bokani}, J.~{Hassan}, and S.~S. {Kanhere}, ``{AETD: An
  Application Aware, Energy Efficient Trajectory Design for Flying Base
  Stations},'' in \emph{2019 IEEE 14th Malaysia International Conference on
  Communication (MICC)}, December 2019.

\bibitem{sutton1998reinforcement}
R.~S. Sutton and A.~G. Barto, ``Reinforcement learning: an introduction
  cambridge,'' \emph{MA: MIT Press.[Google Scholar]}, 1998.

\bibitem{anschel2017averaged}
O.~Anschel, N.~Baram, and N.~Shimkin, ``Averaged-dqn: Variance reduction and
  stabilization for deep reinforcement learning,'' in \emph{International
  Conference on Machine Learning}, 2017, pp. 176--185.

\end{thebibliography}
\end{document}